\documentclass[aps,prl,twocolumn,groupedaddress,amsmath,amssymb]{revtex4-1}
\usepackage{graphicx}  
\usepackage{dcolumn}   
\usepackage{bm}        
\usepackage{verbatim}   
\usepackage{amssymb}
\usepackage{mathtools}
\usepackage{esint}
\usepackage{amsfonts}
\usepackage{amsmath}
\usepackage{graphicx}
\usepackage{bm}
\usepackage{subfig,tikz}
\usepackage{wrapfig}
\usepackage{xspace}
\usepackage{enumerate}
\usepackage{amsthm}

\newtheorem*{theorem*}{Theorem}

\newcommand{\R}{\mathbb R}

\newcommand{\tr}{\operatorname{Tr}}

\begin{document}
\title[Cosmic Cloaking of Rich Extra Dimensions]{Cosmic Cloaking of Rich Extra Dimensions}


\author{Aghil Alaee}
\address{Department of Mathematics, Clark University, Worcester, MA 01610, USA,\\ Center of Mathematical Sciences and Applications, Harvard University, Cambridge, MA 02138, USA}
\email{aalaeekhangha@clarku.edu}

\author{Marcus Khuri}
\address{Department of Mathematics, Stony Brook University, Stony Brook, NY 11794, USA}
\email{khuri@math.sunysb.edu}

\author{Hari Kunduri}
\address{Department of Mathematics and Statistics, Memorial University of Newfoundland, St John's NL A1C 4P5, Canada}
\email{hkkunduri@mun.ca}


\begin{abstract}  \noindent
We present arguments that show why it is difficult to see \emph{rich} extra dimensions in the Universe. More precisely, we study the conditions under which significant size and variation of the extra dimensions in a Kaluza-Klein compactification lead to a black hole in the lower dimensional theory.
The idea is based on the hoop (or trapped surface) conjecture concerning black hole existence, as well as on the observation that dimensional reduction on macroscopically large, twisted, or highly dynamical extra dimensions contributes positively to the energy density in the lower dimensional theory and can induce gravitational collapse.
We analyze these conditions and find that in an idealized scenario a threshold for the size exists, on the order of $10^{-19}m$, such that extra dimensions of length above this level must lie inside black holes, thus shielding them from the view of outside observers.  The threshold is highly dependent on the size of the Universe, leading to the speculation that in the early stages of evolution truly macroscopic and large extra dimensions would have been visible.
\end{abstract}

\maketitle

The idea that there exist extra spatial dimensions beyond the familiar three has persisted for a century.  It cannot be dismissed outright as mere mathematical speculation, as it supplies plausible solutions to difficult problems from particle physics to cosmology. Excitations in extra small dimensions lead to massive particles that could be a candidate for dark matter \cite{darkm}, or help explain the small mass of neutrinos \cite{neutrino}.  The idea has also led to a number of insights in quantum gravity, such as the gauge theory-gravity correspondence~\cite{Aharony:1999ti}, and black hole entropy counting~\cite{Strominger:1996sh}.  Macroscopically `large' extra dimensions have remarkably been suggested as an elegant solution to the hierarchy problem~\cite{ArkaniHamed:1998rs,  Antoniadis:1998ig, ArkaniHamed:1998nn, Randall:1999ee}. The issue of detecting these dimensions in the current generation of experiments is highly model-dependent \cite{Aaboud:2017yvp, Aaboud:2017phn,Aaboud:2017yyg}, but should they exist, it would be reasonable to expect their characteristic length to be no larger than $\sim 10^{-18}m$, as the LHC has probed length scales down to an attometre \cite{Hoecker}. 
This apparent smallness of extra dimensions remains to be satisfactorily explained.

In this note we propose a possible resolution: if there are regions of 4-dimensional spacetime in which extra compact dimensions are sufficiently \emph{rich} to be observed, such regions must be trapped behind black holes.  We will define carefully below what this means.  Our argument produces a rough estimate on the threshold geometry at which horizon formation would occur.  The underlying mechanism is that the size, and the amount that extra dimensions warp and twist over a given spacetime region, produces an effective energy-momentum density which in turn can lead to collapse. We argue that once the extra dimensions get sufficiently large, given a fixed amount of twist, they must be hidden behind a black hole.

The intuitive model of black hole formation is that if matter is enclosed in a sufficiently small region,  a self-gravitating system should collapse with a horizon arising to enclose it.  This heuristic picture is expressed in terms of the \emph{hoop conjecture} \cite{Hoop}.  A refined form of this idea is the \emph{trapped surface conjecture} \cite{Seifert}, which asserts that a trapped surface forms in the vicinity of a body $U$ if the following inequality holds
\begin{equation}\label{TSC}
\mathcal{R}(U) \lesssim \frac{G}{c^2}\textbf{m}(U),
\end{equation}
where $\mathbf{m}(U)$ is a measure of the mass and $\mathcal{R}(U)$ expresses (in units of length) the size of $U$. The symbol $\lesssim$ indicates that there is a universal constant involved that depends on the definition of $\mathcal{R}$ and $\mathbf{m}$.  The trapped surface conjecture has been rigorously established in spherical symmetry \cite{Khuri, MM} with $\mathcal{R}(U)$ given by the radius, and without symmetry hypotheses \cite{SY1} where $\mathcal{R}(U)$ is determined from the largest embedded torus within $U$.

An inequality of the form \eqref{TSC} is precisely stated in terms of the initial value formulation of general relativity, which is the proper setting for studying dynamics.  When \eqref{TSC} holds, it can be shown that an apparent horizon must be present within the initial data. The existence of an apparent horizon signals that a black hole must be contained in the spacetime. In fact,
once a trapped surface or apparent horizon is detected, the Hawking-Penrose singularity theorems \cite{HawkingEllis} together with weak cosmic censorship \cite{Penrose} imply that a horizon will form. The advantage of this approach is that the formation of black holes can be detected without knowledge of the full evolution.

We will demonstrate our proposal in the setting of standard Kaluza-Klein theory, in which there is a single extra spatial circle direction `twisted' over the usual 4-dimensional spacetime.  The ambient 5-dimensional spacetime is taken to satisfy the vacuum Einstein equations.  The length $L$ of this circle, and the amount that it twists over the 4-dimensional spacetime, vary as one moves along this base space.  Roughly, when the length of the circle or the amount that it twists, is too large or changes in a sufficiently rapid fashion over a domain of fixed size, a trapped region must form.  We will present precise conditions on these geometric quantities that produce apparent horizons, and argue that therefore such a \textit{rich} extra dimension would be inaccessible to experimental detection.

Consider the standard warped product with $U(1)$ bundle fibration, in which
the vacuum 5-dimensional spacetime metric $\mathrm{g}_5$ takes the form
\begin{equation}\label{5d}
\mathrm{g}_5 = e^{\frac{2}{\sqrt{3}} \phi} \mathbf{g} + e^{-\frac{4}{\sqrt{3}} \phi} \left(dz + 2 \mathbf{A}\right)^2 ,
\end{equation}
where $\mathbf{g}$ represents the 4-dimensional spacetime metric. The $z$ coordinate parameterizes the $S^1$ direction and is understood to be periodically identified, $z \sim z + 2\pi \ell$, where $\ell$ is a canonical length scale which we may choose to be a meter. $\mathbf{A}$ is a 1-form on spacetime which can be interpreted as measuring how the extra $S^1$ dimension is twisted above the base spacetime, and geometrically determines the connection of the $U(1)$ bundle.  
The circumference of the $S^1$ direction is $2\pi L$ where
\begin{equation}\label{L}
L = \sqrt{\mathrm{g}_{zz}} \ell = e^{-\frac{2}{\sqrt{3}} \phi} \ell .
\end{equation}

Consider now the 4-dimensional theory with spacetime $(N,\mathbf{g})$, 2-form field strength $F = d\mathbf{A}$, and scalar field (dilaton) $\phi$ described by the action
\begin{equation}\label{action}
S = \frac{1}{2\kappa}\int_{N} \left({R}(\mathbf{g}) - 2 |d \phi|^2_{\mathbf{g}} - e^{-\alpha \phi} |F|^2_{\mathbf{g}} \right)d\text{vol}_{\mathbf{g}},
\end{equation}
where $\kappa =\frac {8\pi G}{c^4}$, $R(\mathbf{g})$ and $d\text{vol}_{\mathbf{g}}$ are the scalar curvature and volume form of $\mathbf{g}$ respectively.  For $\alpha = 2\sqrt{3}$, the field equations arising from this action are equivalent to the 5-dimensional vacuum Einstein equations for the metric $\mathrm{g}_5$. We are primarily interested in this case, but for generality we will leave it unfixed.  Note that for $\alpha=0$ and $\phi \equiv 0$, the action \eqref{action} reduces to Einstein-Maxwell theory whereas for $\alpha =2$, it corresponds to a part of the low energy effective action of string theory. We treat $F$ as the curvature of the $U(1)$ connection $\mathbf{A}$ which invariantly measures how the $S^1$ fibers twist over $N$. The field equations are
\begin{equation} \label{FE}
\begin{aligned}
\Delta_{\mathbf{g}} \phi =& -\frac{\alpha}{4} e^{-\alpha \phi} |F|^2_{\mathbf{g}} ,\qquad \text{div}_{\mathbf{g}} \left(e^{-\alpha \phi}  F\right)=0,\\
\mathbf{R}_{ab}=& 2 \bm{\nabla}_a\phi\bm{\nabla}_b\phi +2\mathcal{L}_{ab}, \\
\end{aligned}
\end{equation}
where $\bm{\nabla}$ and $\mathbf{R}_{ab}$ are
the connection and Ricci curvature with respect to $\mathbf{g}$, and
\begin{equation}
\mathcal{L}_{ab}=e^{-\alpha \phi} \left ( \mathbf{g}^{cd} F_{ac} F_{bd} - \frac{1}{4} \mathbf{g}_{ab}|F|^2_{\mathbf{g}} \right).
\end{equation}
Furthermore, the effective 4-dimensional stress-energy tensor induced from the 5-dimensional vacuum equations is given by
\begin{equation}
\begin{split}
\kappa T_{ab} &=2 \bm{\nabla}_a \phi \bm{\nabla}_b\phi - |\bm{\nabla}\phi|_{\mathbf{g}}^2 \mathbf{g}_{ab}+ 2 \mathcal{L}_{ab}.
\end{split}
\end{equation}

Let $M$ be a compact spacelike hypersurface in $N$ with unit timelike normal vector field $n$, an induced positive definite metric $g$ and extrinsic curvature $k$, as well as induced `electric' and `magnetic' spatial vector fields $E=F(n,\cdot)$ and $B=\star F(n,\cdot)$.  An initial data set $(M, g, k, E, B, \phi)$ for the Einstein-Maxwell-scalar field system \eqref{FE} must satisfy the following constraint equations
\begin{equation}\label{eq2}
\begin{aligned}
2\kappa  \mu=&R(g) + (\tr k)^2 -|k|^2_g,\quad \text{div}_g B =0,   \\
\kappa J=&\text{div}_g (k - (\tr k) g),\quad \text{div}_g E = \alpha g\left(E,\nabla \phi\right),
\end{aligned}
\end{equation}
where $\mu$ and $J$ are energy and momentum densities, respectively, defined by
\begin{equation}
\begin{split}
\kappa \mu &=(n \cdot d\phi)^2 + |\nabla\phi|^2_g + e^{-\alpha \phi} \left( |E|^2_g + |B|^2_g \right),\\
 \kappa J&=2 (n \cdot d\phi)\nabla \phi  - 2e^{-\alpha \phi} E\times B.
\end{split}
\end{equation}
The function $n \cdot d\phi$ represents, for the initial value problem, the freely specifiable initial velocity of $\phi$. Define \emph{characteristic constants $\mathcal{E}_i$} of the initial data set $(M, g, k, E, B, \phi)$ by
\begin{equation}\label{Ei}
\begin{split}
\mathcal{E}^3_1&=\fint_{M}\ell^5\,e^{-\alpha\phi}\left(|E|_g-|B|_g\right)^2\,d\text{vol}_g,\\
\mathcal{E}^3_2&=\fint_{M}\ell^5\,\left(|n\cdot d\phi|-|\nabla\phi|_g\right)^2\,d\text{vol}_g,
\end{split}
\end{equation}
where the `slashed' integral indicates an average over $M$, and the quantities $\mathcal{E}_i$ are normalized to have units of length. We then arrive at the following black hole existence result, where the size of the spacelike hypersurface is measured in terms of radius and volume.

\begin{theorem*}\label{thm1}
If at least one of the two characteristic constants of an initial data set $(M, g, k, E, B, \phi)$ for the Einstein-Maxwell-scalar field system \eqref{FE} satisfies the \emph{richness} condition
\begin{equation}\label{richcond}
\mathcal{E}_i^3\gtrsim \ell^5\,
\frac{\mathrm{Rad}(M)}{\mathrm{Vol}(M)},\quad i=1\text{  or }\text{ }2,
\end{equation}
then there exists an apparent horizon within $M$.
\end{theorem*}

The proof of this result is by contradiction, and follows from the trapped surface conjecture. According to equation \eqref{TSC} and using the following definition of mass
\begin{equation}\label{masschoice}
\mathbf{m}(M)=\frac{1}{c^2}\int_{M}(\mu-|J|_g)\,d\text{vol}_g,
\end{equation}
we conclude that the contrapositive of the trapped surface conjecture can be expressed as follows: if the initial data set is devoid of apparent horizons then
\begin{equation}\label{eqhoop}
\mathrm{Rad}(M)\gtrsim \kappa \int_{M}(\mu-|J|_g)\,d\text{vol}_g.
\end{equation}
Observe that the Cauchy-Schwarz inequality yields
\begin{align}\label{ine1}
\begin{split}
\kappa \left(\mu-|J|_g\right)&\geq e^{-\alpha\phi}\left(|E|_g-|B|_g\right)^2\\
&\quad+\left(|n\cdot d\phi|-|\nabla\phi|_g\right)^2 .
\end{split}
\end{align}
Dividing \eqref{eqhoop} by $\text{Vol}(M)$ and combining with \eqref{ine1} then produces
\begin{equation}
\mathcal{E}_i^3\leq\mathcal{E}_1^3+\mathcal{E}_2^3\lesssim \ell^5\,\frac{\text{Rad}(M)}{\text{Vol}(M)},
\end{equation}
for $i=1,2$. Therefore, the reverse of this inequality implies that there must exist an apparent horizon within $M$.

Schoen and Yau have established a rigorous mathematical formulation and proof of the result that sufficient concentration of matter leads to horizon formation~\cite{SY1}, a result that naturally arose from their arguments establishing the positive mass theorem \cite{SchoenYauII}. Our choice of mass in \eqref{masschoice}, is motivated by their usage of the quantity $\mu-|J|_g$ which is related to the dominant energy condition. An analogous statement to the theorem may be proved, following the methods of \cite{SY1}, with the main difference that the characteristic constants $\mathcal{E}_i$ are replaced with weighted integrals involving the principal eigenfunction of a certain differential operator on $M$. We also point out, that the definition of mass \eqref{masschoice} and usage of the radius to measure size have appeared in other formulations of the trapped surface conjecture, see for example \cite{Khuri,MM}. Furthermore, it should be pointed out \cite{PaetzSimon} that the 2-dimensional apparent horizons discussed here, correspond to 3-dimensional apparent horizons in the 5-dimensional spacetime via a quotient by the $U(1)$ action.

Let us now examine some consequences of the theorem. In particular, we will consider the case in which the initial data set encompasses the entire universe. The radius of the observable universe has been observed to be approximately $10^{26} m$ \cite{Halpern}. According to the most recent analysis of the cosmic microwave background \cite{Planck}, the universe does not exhibit any known topological features, and thus its time slices may be approximated by the simply connected constant curvature model of a round 3-sphere (see also \cite{GallowayKhuriWoolgar}), or flat Euclidean 3-space in which case $M$ is taken to be a large ball. We are thus able to compute the ratio of radius to volume appearing in the theorem, namely
\begin{equation}
\left(\ell^5\frac{\text{Rad}(M)}{\text{Vol}(M)}\right)^{1/3}\sim 10^{-19}m.
\end{equation}
It follows that if $\mathcal{E}_i\gtrsim 10^{-19}m$ for either $i=1$ or $2$, then a black hole must form due to  concentration of the geometry, or richness, of the extra dimension. Although the theorem does not determine exactly where in $M$ the apparent horizon is located, it is reasonable to surmise that the regions where the concentration is highest should be where the black holes form.

Next we examine the individual cases in which each characteristic constant satisfies the richness condition \eqref{richcond}. Observe that Jensen's inequality implies a lower bound for the first characteristic constant in terms of the extra dimension's size and twist
\begin{equation}
\mathcal{E}_1 \geq \fint_{M} L \cdot\ell^{2/3}\left(|E|_g-|B|_g\right)^{2/3}\,d\text{vol}_g.
\end{equation}
Hence, if the twisting is generically on the order of $\left(|E|_g-|B|_g\right)^{2/3}\sim 10^{-p}\ell^{-2/3}$, then the richness condition will be satisfied when, on the scale of the universe, the average circumference of the extra dimension satisfies $\mathrm{avg}(L)\gtrsim 10^{-19+p}m$. We then have the existence of black holes due to the excessive average size of the extra dimension, and as argued the horizons should be located in regions where this size is greatest. In this way, large extra dimensions are hidden from the view of outside observers. On the other hand, if the size is generically $L\sim 10^{-p}m$, then the richness condition is satisfied when the average twisting surpasses the threshold $10^{-19+p}\ell^{-1}$, in which case sufficiently twisted extra dimensions are hidden behind horizons.

Consider now the second characteristic constant, which concerns a measurement of rate of change in time and space of the size of the circle fibers
\begin{equation}
\mathcal{E}_2 \gtrsim\fint_{M}\ell^{5/3} \left(|n\cdot d\log L| -|\nabla\log L|\right)^{2/3}\, d\text{vol}_g.
\end{equation}
The richness condition will be satisfied if on average, throughout a time slice of the Universe, the rate of change of the extra dimension's size is greater than $10^{-19}m$. Thus, highly dynamical extra dimensions or those with extreme spatial size variations are enclosed inside black holes. This may be compared to the result discussed in
\cite{Draper}, where localized variations of a light scalar field were observed to induce collapse. In a different direction, Penrose \cite[Section 31.12]{Penrose1} has sketched an instability argument that suggests the emergence of singularities in supersymmetric compactifications due to Planck-sized extra dimensions.

It should be pointed out that the threshold of $10^{-19}m$ is tied to the diameter of the universe in the current epoch.
Thus, we speculate that at earlier times, when the diameter was significantly smaller the threshold would be much larger. In fact, in the early stages of the universe truly macroscopic and even large extra dimensions would have been visible, and allowed to change rapidly without being trapped behind horizons. Conversely, as the universe expands into the future, the threshold will drop and eventually achieve a level on par with the Planck length, making it virtually impossible to detect the extra dimensions.

The conclusions of this note suggest that there is a fundamental tension between the `richness' - the  twisting and warping - of extra dimensions and the ability to explore these dimensions experimentally.  We have restricted attention to the simplest model of extra dimensions to emphasize key features of the arguments. It should be the case, however, that additional (possibly curved) spatial dimensions should produce further positive contributions to the energy density of the effective theory, leading to similar effects. The same is also true if additional matter fields are present and satisfy the dominant energy condition, or if there is a nonnegative cosmological constant.
Thus, we expect that the results demonstrated here can be generalized to more complicated models in a robust manner.

{\bf Acknowledgments.}
The authors would like to thank G. Horowitz, H. Reall, M. Rocek, W. Simon, J. Smillie, and C. Vafa for helpful comments.

A. Alaee acknowledges the support of an AMS Simons Travel Grant. M. Khuri acknowledges the support of NSF Grant DMS-1708798, and Simons Foundation Fellowship 681443. H. Kunduri acknowledges the support of NSERC Grant RGPIN-2018-04887.

\bibliographystyle{apsrev4-1}

\end{document}